\documentclass[english,aps,manuscript]{revtex4-1}
\usepackage[T1]{fontenc}
\usepackage[latin9]{inputenc}
\usepackage[active]{srcltx}
\usepackage{amstext}
\usepackage{amssymb}
\usepackage{graphicx}
\usepackage{esint}

\makeatletter

\DeclareRobustCommand{\greektext}{%
  \fontencoding{LGR}\selectfont\def\encodingdefault{LGR}}
\DeclareRobustCommand{\textgreek}[1]{\leavevmode{\greektext #1}}
\DeclareFontEncoding{LGR}{}{}
\DeclareTextSymbol{\~}{LGR}{126}
\newcommand{\lyxmathsym}[1]{\ifmmode\begingroup\def\b@ld{bold}
  \text{\ifx\math@version\b@ld\bfseries\fi#1}\endgroup\else#1\fi}

\@ifundefined{textcolor}{}
{%
 \definecolor{BLACK}{gray}{0}
 \definecolor{WHITE}{gray}{1}
 \definecolor{RED}{rgb}{1,0,0}
 \definecolor{GREEN}{rgb}{0,1,0}
 \definecolor{BLUE}{rgb}{0,0,1}
 \definecolor{CYAN}{cmyk}{1,0,0,0}
 \definecolor{MAGENTA}{cmyk}{0,1,0,0}
 \definecolor{YELLOW}{cmyk}{0,0,1,0}
 }

\makeatother

\usepackage{babel}
\begin{document}

\title{Geodesic Structure of the Noncommutative Schwarzschild Anti-de Sitter
Black Hole I: Timelike Geodesics}

\author{Alexis Larrañaga}

\address{National Astronomical Observatory. National University of Colombia.}
\begin{abstract}
By considering particles as smeared objects, we investigate the effects
of space noncommutativity on the geodesic structure in Schwarzschild-AdS
spacetime. By means of a detailed analysis of the corresponding effective
potentials for particles, we find the possible motions which are allowed
by the energy levels. Radial and non-radial trajectories are treated
and the effects of space noncommutativity on the value of the precession
of the perihelion are estimated. We show that the geodesic structure
of this black hole presents new types of motion not allowed by the
Schwarzschild spacetime. 


\end{abstract}
\maketitle

\section{Introduction}

The presence of a vacuum energy (cosmological constant) in theoretical
models has been considered in relation to unification, such as superstring
theory, and to cosmology and astrophysics. This has motivated consideration
of spherical symmetric spacetimes with non-zero vacuum energy in order
to study the well-known effects predicted by general relativity for
planetary orbits and massless particles. This study implies the determination
of the geodesic structure of spacetimes \cite{kottler}. Timelike
geodesics for a positive cosmological constant were investigated in
\cite{key-2} using an effective potential method to find the conditions
for the existence of bound orbits. The analysis of the effective potential
for radial null geodesic in Reissner\textendash{}Nordstrom\textendash{}deSitter
and Kerr\textendash{}deSitter spacetime was realized in \cite{key-4}
and \cite{key-5}. Podolsky \cite{key-6} investigated all possible
geodesic motions for extreme Schwarzschild\textendash{}de Sitter spacetime.
Finally, Cruz et.al. \cite{key-7} made a complete disussion on the
geodesic structure of Schwarzschild-anti de Sitter (Schw-AdS) black
hole.

On the other hand, gedanken experiments that aim at probing spacetime
structure at very small distances support the idea that noncommutativity
of spacetime is a feature of Planck scale physics. It appears to happen
that due to gravitational back reaction, one cannot test spacetime
at Planck scale. Its description as a smooth manifold becomes therefore
a mathematical assumption no more justified by physics and therefore,
it is natural to relax this assumption and conceive a more general
noncommutative spacetime, where uncertainty relations and discretization
naturally arise. 

As is well known, noncommutativity is the central mathematical concept
expressing uncertainty in quantum mechanics, where it applies to any
pair of conjugate variables, such as position and momentum. Thus,
one can easily imagine that position measurements might fail to commute
and this fact will be described using noncommutativity of spacetime
coordinates. The noncommutativity of spacetime coordinates can be
encoded in the commutator \cite{NC1,NC2,NC3,NC4,NC5,NC6,NC7,NC8}

\begin{equation}
\left[x^{\mu},x^{\nu}\right]=i\varepsilon^{\mu\nu}
\end{equation}

where $\varepsilon^{\mu\nu}$ is a real, antisymmetric and constant
tensor, which determines the fundamental cell discretization of spacetime
(in the same way as the Planck constant $\hbar$ discretizes the phase
space). In four dimensions and using an adequate choice of coordinates,
this tensor can be brought to the form

\begin{equation}
\varepsilon^{\mu\nu}=\left(\begin{array}{cccc}
0 & \varepsilon^{2} & 0 & 0\\
-\varepsilon^{2} & 0 & \varepsilon^{2} & 0\\
0 & -\varepsilon^{2} & 0 & \varepsilon^{2}\\
0 & 0 & -\varepsilon^{2} & 0
\end{array}\right),
\end{equation}
where $\varepsilon$ is a constant with dimension of length.

The modifications induced by noncommutativity on the classical orbits
of particles in a central force potential has been considered by Benczik
et al \cite{NCg1}, by Mirza and Dehghani \cite{NCg2} and by Romero
and Vergara \cite{NCg3}. These investigations let them impose a constraint
on the minimal observable length and noncommutativity parameter in
comparison with observational data of Mercury. The stability of planetary
orbits of particles in noncommutative space has been studied both
in central force and Schwarzschild background by Nozari and Akhshabi
\cite{NCg4} and the Kepler problem in noncommutative Schwarzschild
geometry in \cite{NCg5}. The purpose of this paper is to investigate
these effects on the orbits of a test particle in noncommutative Schwarzschild-AdS
(NCSchw-AdS) geometry to generalize the geodesic structure studied
in \cite{key-7}.

\section{The Noncommutative Schwarzschild-AdS Black HOle}

It has been shown that noncommutativity eliminates point-like structures
in favor of smeared objects in flat spacetime \cite{NCBH0,NCBH1}.
The effect of smearing can be mathematically implemented as a substitution
rule: position Dirac-delta function can be replaced everywhere with
a Gaussian distribution of minimal width $\varepsilon$. In this framework,
the mass density of a static, spherically symmetric, smeared, particle-like
gravitational source can be shown by a Gaussian profile \cite{NCBH,NCBH2,NCBH3,NCBH4}.
Solving the Einstein field equations, one can find the metric for
a static spherically symmetric object with total mass $M$ in a noncommutative
spacetime with negative cosmological constant $\lyxmathsym{\textgreek{L}}=-\frac{3}{L^{2}}$
as \cite{NCSadS}

\begin{equation}
ds^{2}=-f\left(r\right)dt^{2}+\frac{dr^{2}}{f\left(r\right)}+r^{2}d\Omega^{2}\label{eq:NCSAdS}
\end{equation}

where the lapse function is

\begin{equation}
f\left(r\right)=1-\frac{4M\gamma\left(\frac{3}{2};\frac{r^{2}}{4\varepsilon^{2}}\right)}{r\sqrt{\pi}}+\frac{r^{2}}{L^{2}}
\end{equation}

and $\gamma$ is the lower incomplete gamma function,

\begin{equation}
\gamma\left(\frac{3}{2};x\right)=\int_{0}^{x}dtt^{1/2}e^{-t}.
\end{equation}

The horizon equation $f\left(r_{+}\right)=0$ depends on two parameters,
$M$ and $L$ and cannot be solved in a closed form. However, we can
draw plots to study the occurrence of horizons. In order to do it,
we will write the lapse function as 

\begin{equation}
f\left(x\right)=1-\frac{4m\gamma\left(\frac{3}{2};\frac{x^{2}}{q^{2}}\right)}{x\sqrt{\pi}}+x^{2}
\end{equation}

where we have defined $x=\frac{r}{L}$, $m=\frac{MG}{L}$ and $q=\frac{2\varepsilon}{L}$.
In Figure 1 the plot of $f\left(x\right)$ show that all curves start
at $f\left(0\right)=1$, indicating that the spacetime is regular
and therefore geodesically complete. The behavior of the curves shows
three possibilities,
\begin{enumerate}
\item For $m>m_{0}$ there are two horizons, $x_{-}$ and $x_{+}$ (i.e.
$r_{-}$ and $r_{+}$ )
\item For $m=m_{0}$ there is one degenerate horizon, $x_{0}$ (i.e. $r_{0}$
)
\item For $m<m_{0}$ there is no horizon.
\end{enumerate}
The critical value $m_{0}$ depends on $\varepsilon$ and $L$ and
is determined by the conditions

\begin{equation}
f\left(x\right)=\frac{\partial f}{\partial x}=0.
\end{equation}

\begin{figure}
\includegraphics[scale=0.25]{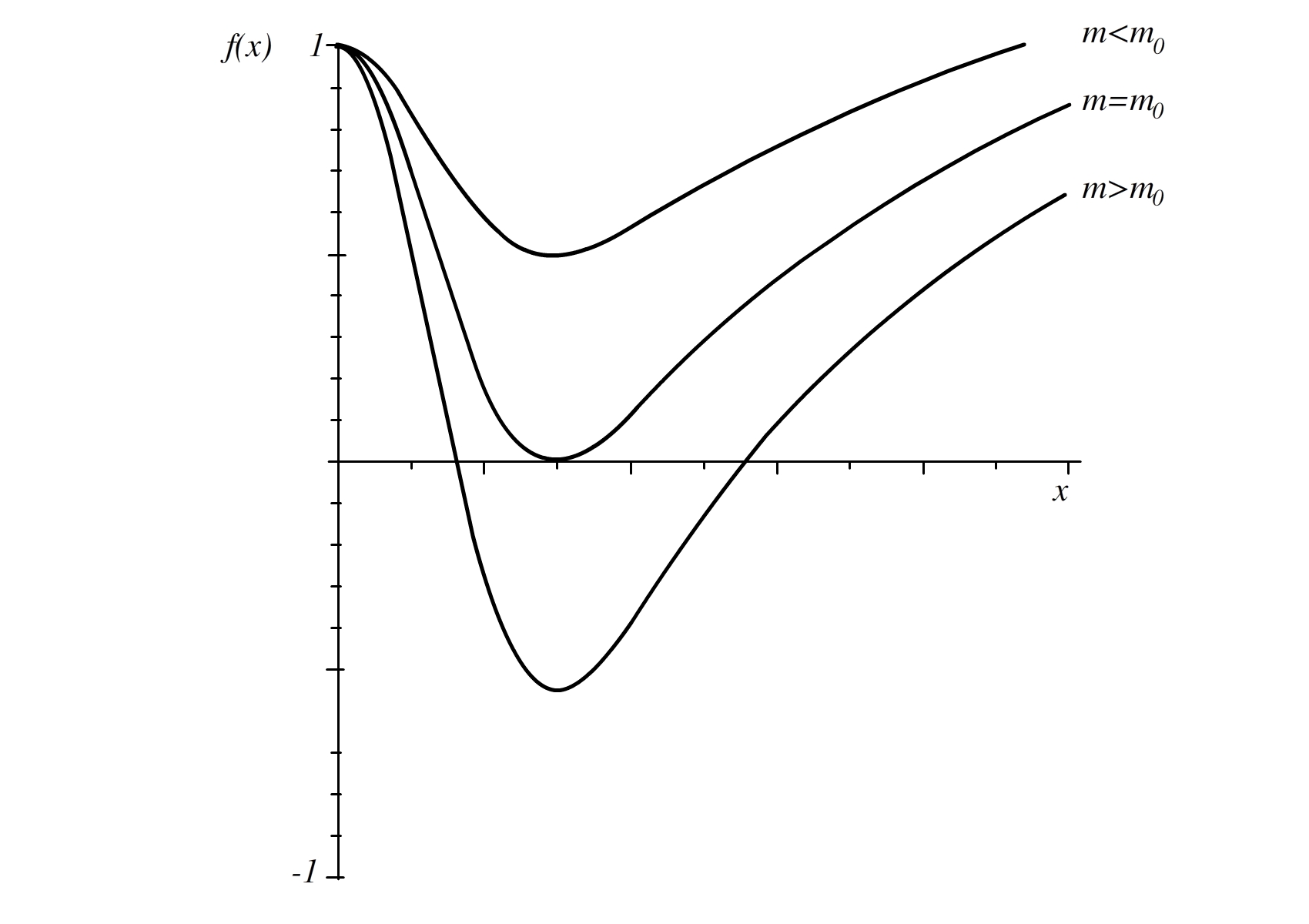}

\caption{$f\left(x\right)$ for different values of $m$. We notice that there
exist three cases, namely two horizons, no horizon and one single
degenerate horizon.}
\end{figure}

\section{TimeLike Geodesics}

In order to find the geodesics structure of the spacetime described
by (\ref{eq:NCSAdS}), we solve the Euler-Lagrange equations for the
variational problem associated to this metric \cite{adler}. The Lagrangian
is

\begin{equation}
\mathcal{L}=-f\left(r\right)\dot{t}^{2}+\frac{\dot{r}^{2}}{f\left(r\right)}+r^{2}\dot{\theta}^{2}+r^{2}\sin^{2}\theta\dot{\varphi}^{2}
\end{equation}

where the dots represent the derivative with respect to the affine
parameter $\tau$, along the geodesic. The equations of motion are

\begin{equation}
\dot{\pi}_{q}=\frac{d}{dt}\left(\frac{\partial\mathcal{L}}{\partial\dot{q}}\right)=\frac{\partial\mathcal{L}}{\partial q}.
\end{equation}

Since $\mathcal{L}$ is independient of $t$ and $\varphi$ there
are two conserved quantities,

\begin{equation}
E=-\frac{\pi_{t}}{2}=f\left(r\right)\dot{t}
\end{equation}

and 

\begin{equation}
\ell=\frac{\pi_{\varphi}}{2}=r^{2}\sin^{2}\theta\dot{\varphi}.
\end{equation}

Meanwhile, the equation of motion for $\theta$ gives

\begin{equation}
\frac{d(r^{2}\dot{\theta})}{d\tau}=r^{2}sin\theta cos\theta\dot{\phi}^{2}.
\end{equation}

Therefore, if we choose the initial condition $\theta=\frac{\pi}{2}$
and $\dot{\theta}=0$, the last equation gives $\ddot{\theta}=0$.
This means that the motion is confined to the plane $\theta=\frac{\pi}{2}$
, which is characteristic of central fields. With this election, the
angular momentum is

\begin{equation}
\ell=r^{2}\dot{\varphi}\label{eq:momentoangular}
\end{equation}

and the Lagrangian becomes

\begin{equation}
\mathcal{L}=-h=-\frac{E^{2}}{f\left(r\right)}+\frac{\dot{r}^{2}}{f\left(r\right)}+\frac{\ell^{2}}{r^{2}},
\end{equation}

where we shall consider $h=1$ for massive particles and $h=0$ for
photons. Solving the above equation for $\dot{r}^{2}$ we obtain the
radial equation which allow us to characterize possible movements
of test particles without and explicit solution of the equation of
motion in the invariant plane. This is

\begin{equation}
\dot{r}^{2}=E^{2}-f\left(r\right)\left(h+\frac{\ell^{2}}{r^{2}}\right)
\end{equation}

or better

\begin{equation}
\dot{r}^{2}=E^{2}-V_{eff}^{2}\label{eq:velocity}
\end{equation}

with the effective potential 
\begin{equation}
V_{eff}^{2}\left(r\right)=\left(1-\frac{4M\gamma\left(\frac{3}{2};\frac{r^{2}}{4\varepsilon^{2}}\right)}{r\sqrt{\pi}}+\frac{r^{2}}{L^{2}}\right)\left(h+\frac{\ell^{2}}{r^{2}}\right).
\end{equation}

For timelike geodesics, $h=1$, the effective potential becomes

\begin{equation}
V_{eff}^{2}\left(r\right)=\left(1-\frac{4M\gamma\left(\frac{3}{2};\frac{r^{2}}{4\varepsilon^{2}}\right)}{r\sqrt{\pi}}+\frac{r^{2}}{L^{2}}\right)\left(1+\frac{\ell^{2}}{r^{2}}\right).
\end{equation}

$V_{eff}^{2}\left(r\right)$ let us solve the equation of motion for
two interesting special cases of massive particle orbits, namely radial
motion and bound orbits.

\subsection{Radial Geodesics}

For radial geodesics, $\ell=0$, we have

\begin{equation}
V_{eff}^{2}\left(r\right)=1-\frac{4M\gamma\left(\frac{3}{2};\frac{r^{2}}{4\varepsilon^{2}}\right)}{r\sqrt{\pi}}+\frac{r^{2}}{L^{2}}.
\end{equation}

The behavior of the effective potential is shown in Figure 2. Note
that in Figure 2 (a) (with $L=1$ and $M=1$) the particle always
moves towards $r=0$; but for greater $M$ or greater $L$ (i.e. smaller
$\Lambda$), the function $V_{eff}^{2}$ has a minimum. Therefore,
for certain values of the energy of the particle moving radially,
it can not reach $r=0$ but is repelled once it has approached to
within some finite distance.

\begin{figure}
\begin{centering}
\includegraphics[scale=0.25]{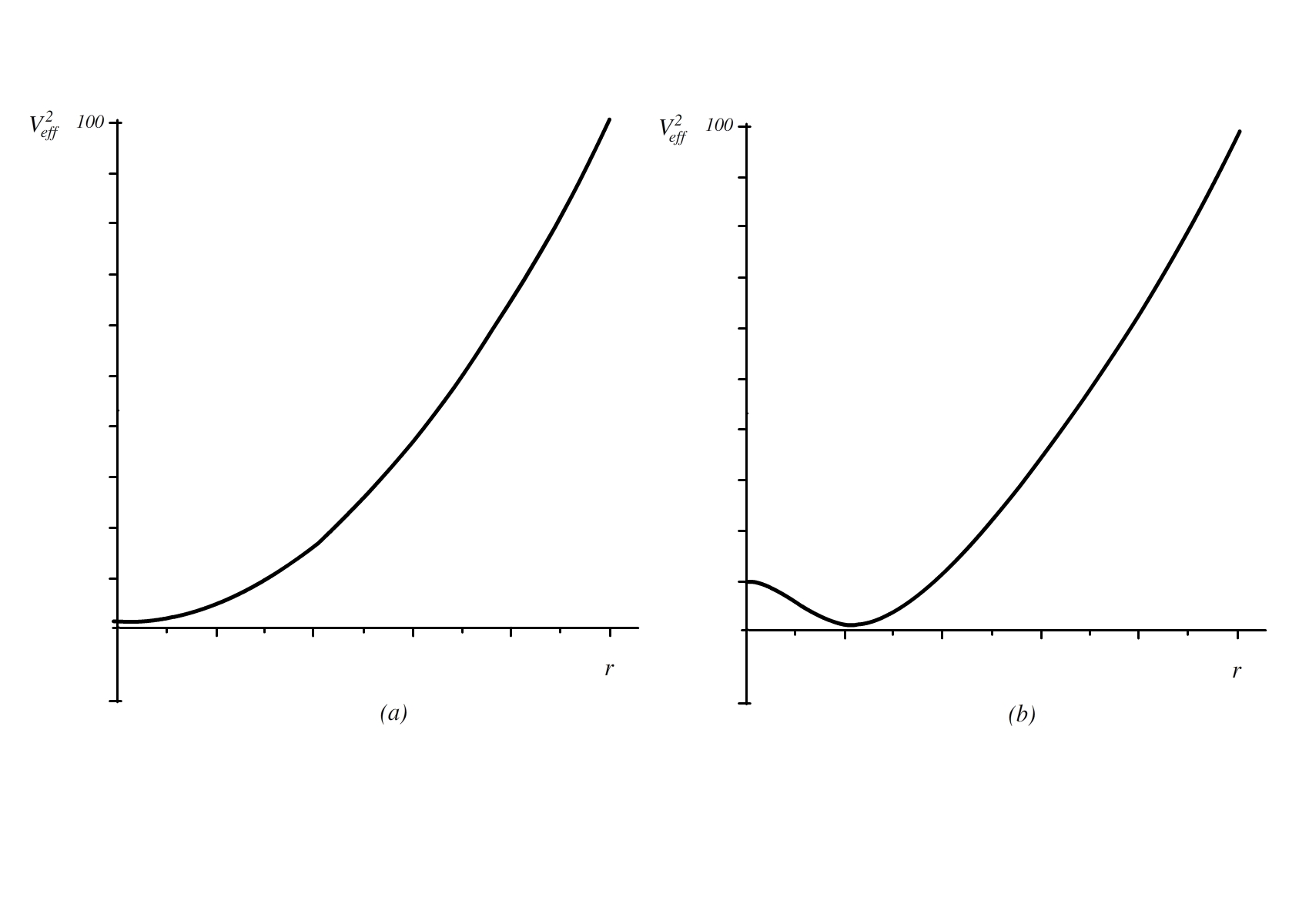}
\par\end{centering}

\caption{The effective potential for radial particles. For figure (a) we use
$\varepsilon=1$ and $L=1$ while figure (b) has $\varepsilon=1$
and $L=10$ in arbitrary units.}

\end{figure}

In the curve representing $V_{eff}^{2}$, particles always plunge
into the horizon from an upper distance determined by the constant
of motion $E$. If the particle is release from rest at a distance
$r=r_{i}$ we have the constant

\begin{equation}
E^{2}=V_{eff}^{2}\left(r_{i}\right)=1-\frac{4M\gamma\left(\frac{3}{2};\frac{r_{i}^{2}}{4\varepsilon^{2}}\right)}{r_{i}\sqrt{\pi}}+\frac{r_{i}^{2}}{L^{2}}
\end{equation}

and the equation of motion can be written as

\begin{equation}
\dot{r}^{2}=\frac{4M}{\sqrt{\pi}}\left[\frac{\gamma\left(\frac{3}{2};\frac{r^{2}}{4l^{2}}\right)}{r}-\frac{\gamma\left(\frac{3}{2};\frac{r_{i}^{2}}{4\varepsilon^{2}}\right)}{r_{i}}\right]-\frac{1}{L^{2}}\left(r^{2}-r_{i}^{2}\right)
\end{equation}
which can be integrated as

\begin{equation}
\tau\left(r\right)=\int_{r_{i}}^{r}\frac{dr}{\sqrt{\frac{4M}{\sqrt{\pi}}\left[\frac{\gamma\left(\frac{3}{2};\frac{r^{2}}{4\varepsilon^{2}}\right)}{r}-\frac{\gamma\left(\frac{3}{2};\frac{r_{i}^{2}}{4\varepsilon^{2}}\right)}{r_{i}}\right]-\frac{1}{L^{2}}\left(r^{2}-r_{i}^{2}\right)}}\label{eq:propertime}
\end{equation}
to give the proper time experienced by a particle in falling from
$r_{i}$ to a coordinate radius $r$. Equation (\ref{eq:propertime})
is plotted in Figure 3 and show that the particle falls towards the
horizon in a finite proper time smaller than that corresponding to
the Schwarzschild-AdS case.

\begin{figure}
\begin{centering}
\includegraphics[scale=0.25]{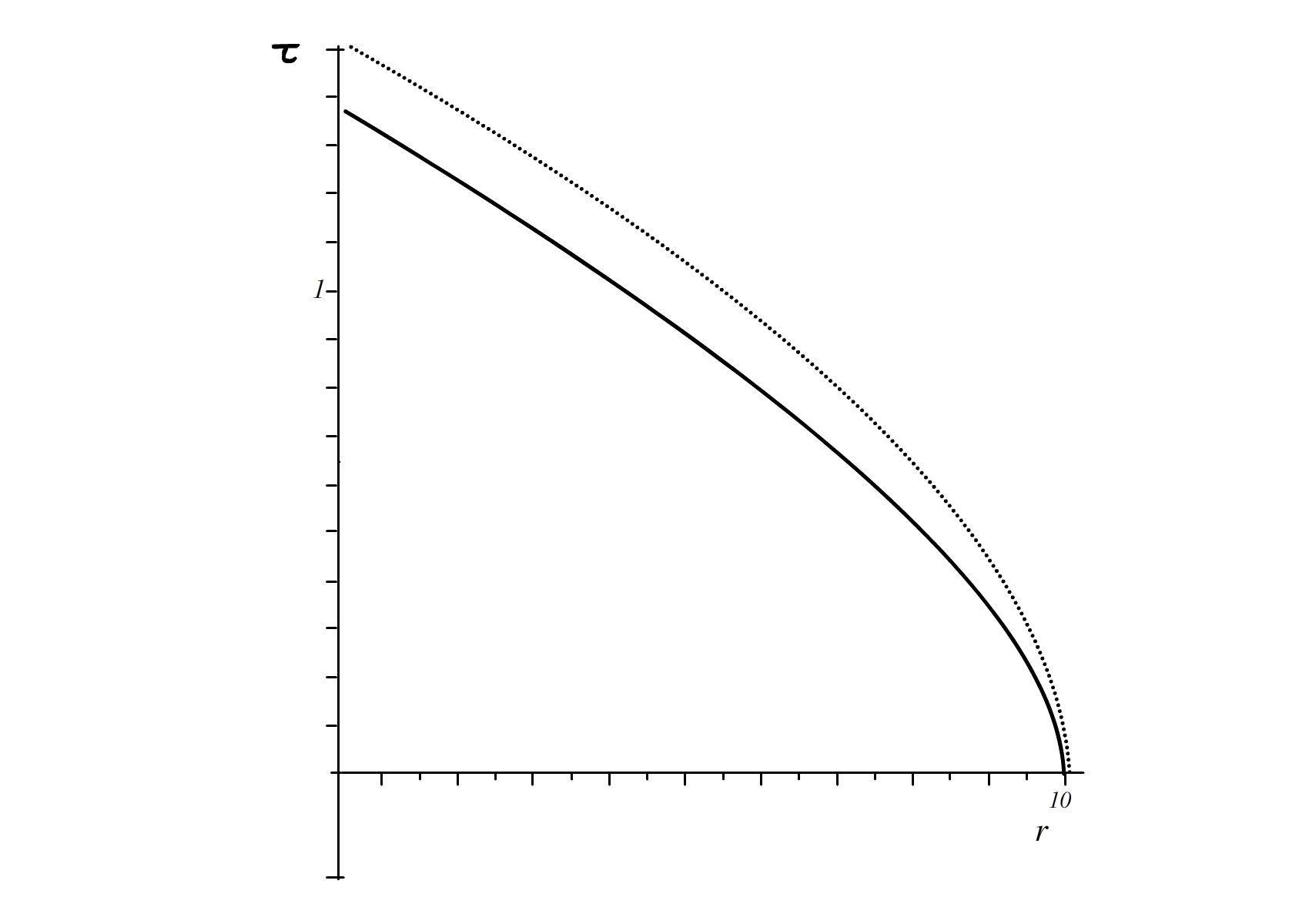}
\par\end{centering}

\caption{Proper time $\tau$ as a function of $r$. The dotted curve corresponds
to the Schwarzschild-AdS metric while the continuous curve is the
noncummutative black hole. In this case $r_{i}=10$.}
\end{figure}

\subsection{The Bound Orbits}

In this case $\ell\neq0$, and

\begin{equation}
V_{eff}^{2}\left(r\right)=\left(1-\frac{4M\gamma\left(\frac{3}{2};\frac{r^{2}}{4\varepsilon^{2}}\right)}{r\sqrt{\pi}}+\frac{r^{2}}{L^{2}}\right)\left(1+\frac{\ell^{2}}{r^{2}}\right).\label{eq:effectivepotential}
\end{equation}

\begin{figure}
\begin{centering}
\includegraphics[scale=0.25]{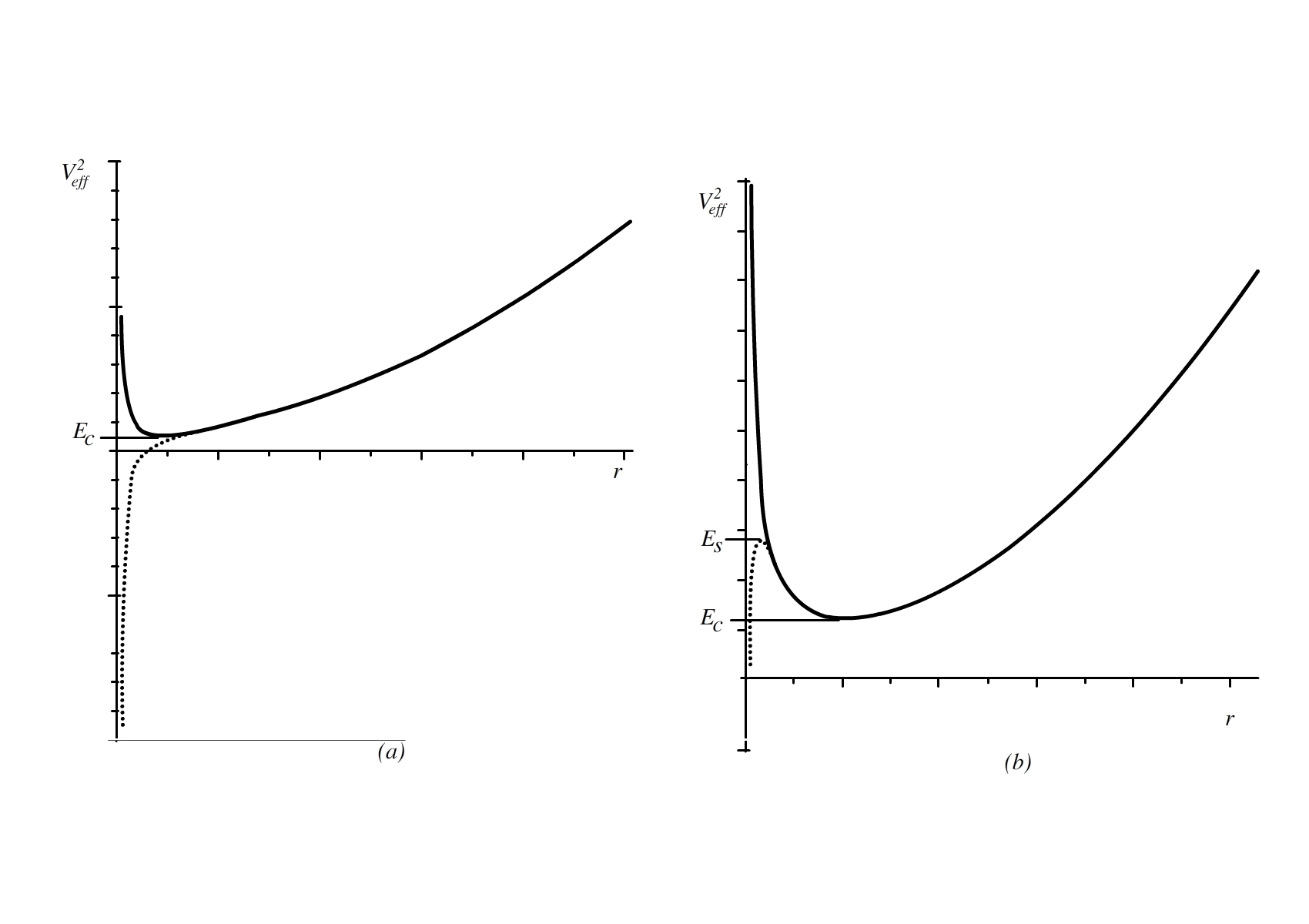}
\par\end{centering}

\caption{The effective potential for non-radial particles in the Schwarzschild-AdS
metric (dotted curve) and non-commutative Schwarzschild-AdS (continuous
curve). In figure (a) we set $L=10$, $l=1$ and $\ell=1$ while in
figure (b) has $L=10$, $l=1$ and $\ell=15$ in arbitrary units.}
\end{figure}

In Figure 4, the effective potential has been plotted for non-radial
particles and compared to the Schwarzschild-AdS case. Note that for
the noncommutative black hole there are always two kinds of allowed
orbits, depending on the value of the constant $E$,
\begin{enumerate}
\item If $E^{2}=E_{c}^{2}$, the particle orbits in a stable circular orbit
at $r=r_{c}$
\item If $E^{2}>E_{c}^{2}$, the particle orbits on a bound orbit in the
range $r_{P}<r<r_{A}$ ($r_{P}$ and $r_{A}$ are the perihelion and
aphelion distances, respectively),
\end{enumerate}
where $E_{c}^{2}$ corresponds to the minimum value of the effective
potential, $E_{c}^{2}=\left.V_{eff}^{2}\right|_{min}$. 

In Figure 4 there is a third possibility in Scharzschild-AdS case
(dotted curve). When $E^{2}=E_{S}^{2}$, the particle can orbit in
an unstable circular orbit. However, this orbit is never allowed for
the noncommutative black hole. The divergency of the continuous curve
in Figure 4 around the origin is a manifestation of the existence
of a minimal lenght scale, which prevents to probe distances smaller
that the fundamental distance $\varepsilon$.

The equation of motion is obtained using equations (\ref{eq:momentoangular})
and (\ref{eq:velocity}) and making the change of variable $u=\frac{1}{r}$,
giving

\begin{equation}
\left(\frac{du}{d\theta}\right)^{2}+u^{2}=\frac{\epsilon^{2}}{\ell^{2}}+\frac{4Mu}{\sqrt{\pi}\ell^{2}}\gamma\left(\frac{3}{2};\frac{1}{4\varepsilon^{2}u^{2}}\right)+\frac{4M}{\sqrt{\pi}}u^{3}\gamma\left(\frac{3}{2};\frac{1}{4\varepsilon^{2}u^{2}}\right)-\frac{1}{L^{2}\ell^{2}u^{2}}
\end{equation}

with

\begin{equation}
\epsilon^{2}=E^{2}-1-\frac{\ell^{2}}{L^{2}}.
\end{equation}

This expression can be rewritten as

\begin{equation}
\left(\frac{du}{d\theta}\right)^{2}=f\left(u\right)\label{eq:aux1}
\end{equation}
with 
\begin{equation}
f\left(u\right)=\frac{\epsilon^{2}}{\ell^{2}}+\frac{4Mu}{\sqrt{\pi}\ell^{2}}\gamma\left(\frac{3}{2};\frac{1}{4\varepsilon^{2}u^{2}}\right)+\frac{4M}{\sqrt{\pi}}u^{3}\gamma\left(\frac{3}{2};\frac{1}{4\varepsilon^{2}u^{2}}\right)-u^{2}-\frac{1}{L^{2}\ell^{2}u^{2}}.
\end{equation}
Considering only orbits which possess perihelia, the point of closest
approach is given by the condition

\begin{equation}
\frac{du}{d\theta}=f\left(u\right)=0
\end{equation}

and for the rest of the orbit $u$ is less than its perihelion value.
Equation (\ref{eq:aux1}) tell us that throughout the orbit $f\left(u\right)\geq0$.
In Figure 5 we show three typical situations in which function $f\left(u\right)$
has one and three zeros. Figure 5b corresponds to an elliptical orbit
with $u$ oscillating in the range $u_{A}<u<u_{P}$ ($u_{P}$ corresponds
to the perihelion while $u_{A}$ is the aphelion). On the other hand,
Figure 5c shows the special case in which $u_{A}=u_{P}=u_{c}$ and
the orbit becomes a circle.

\begin{figure}
\begin{centering}
\includegraphics[scale=0.25]{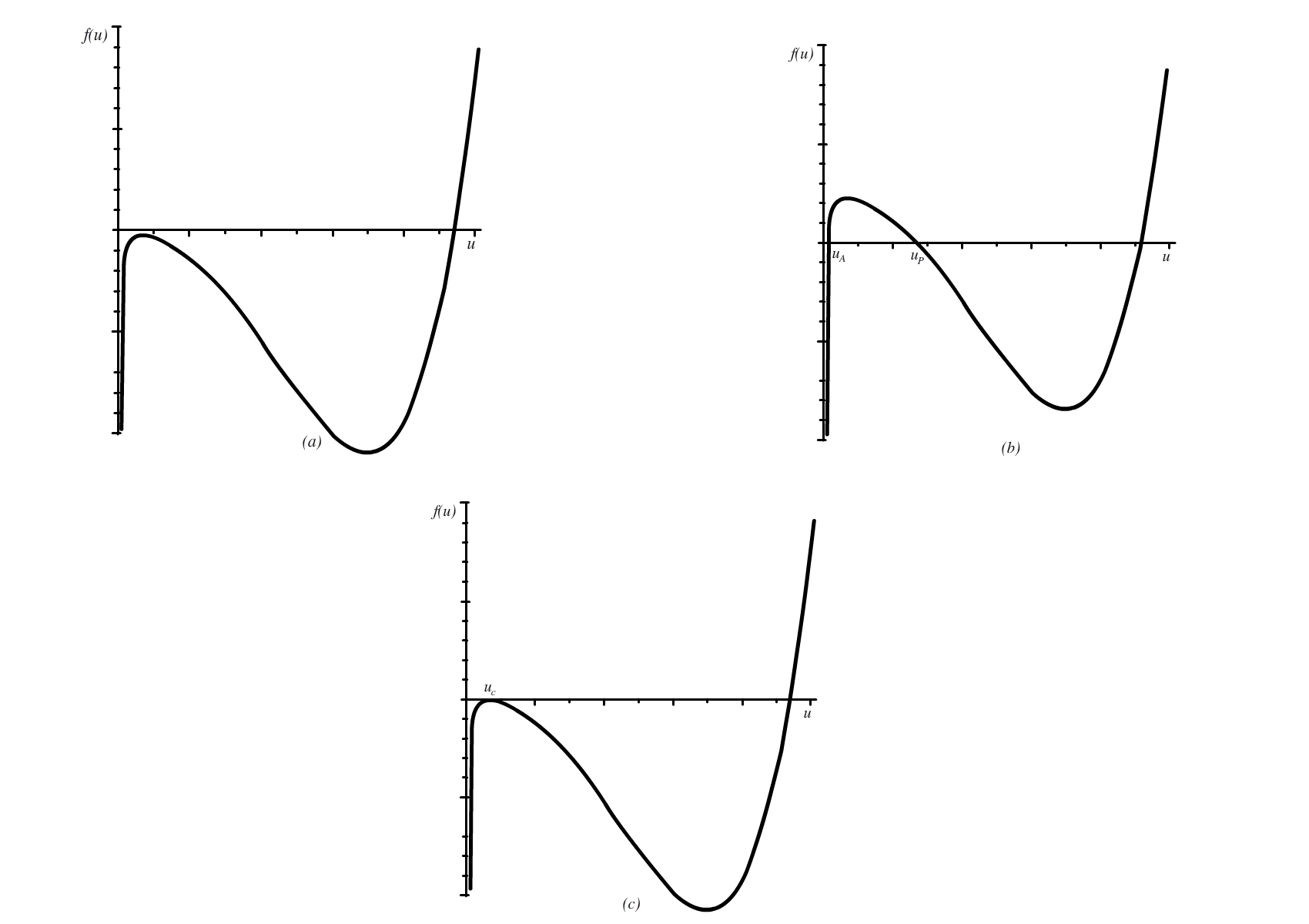}
\par\end{centering}

\caption{Graphs of the function $f\left(u\right)$. In figure (a) we set $L=10$,
$\varepsilon=1$, $\ell=10$ and $\epsilon=0$. Figure (b) has $L=10$,
$\varepsilon=1$, $\ell=10$ and $\epsilon=5$ while Figure (c) use
$L=10$, $\varepsilon=1$, $\ell=10$ and $\epsilon=1.41$}
\end{figure}

We now differentiate this equation with respect to $\theta$, using
the relation 

\begin{equation}
\frac{\partial}{\partial u}\gamma\left(\frac{3}{2},\frac{1}{4\varepsilon^{2}u^{2}}\right)=\frac{e^{-1/4\varepsilon^{2}u^{2}}}{4\varepsilon^{3}u^{4}},\label{eq:auxdif}
\end{equation}
to obtain 

\begin{equation}
\frac{d^{2}u}{d\theta^{2}}+u=\frac{2M}{\sqrt{\pi}\ell^{2}}\gamma\left(\frac{3}{2};\frac{1}{4\varepsilon^{2}u^{2}}\right)+\frac{6M}{\sqrt{\pi}}u^{2}\gamma\left(\frac{3}{2};\frac{1}{4\varepsilon^{2}u^{2}}\right)+\frac{1}{L^{2}\ell^{2}u^{3}}+\frac{M}{2\sqrt{\pi}\ell^{5}}\frac{e^{-1/4\varepsilon^{2}u^{2}}}{u^{3}}+\frac{M}{2\sqrt{\pi}\varepsilon^{3}}\frac{e^{-1/4\varepsilon^{2}u^{2}}}{u}.
\end{equation}

Approximating the incomplete gamma function for long distances to
first order,

\begin{equation}
\gamma\left(\frac{3}{2},\frac{1}{4\varepsilon^{2}u^{2}}\right)\simeq\frac{\sqrt{\pi}}{2}-\frac{1}{2\varepsilon}\frac{e^{-1/4\varepsilon^{2}u^{2}}}{u},\label{eq:aux2}
\end{equation}

the differential equation of the orbit gives

\begin{equation}
\frac{d^{2}u}{d\theta^{2}}+u=\frac{M}{\ell^{2}}+3Mu^{2}+\frac{1}{L^{2}\ell^{2}u^{3}}+\frac{M}{\sqrt{\pi}}e^{-1/4\varepsilon^{2}u^{2}}\left[\frac{3u}{\varepsilon}+\frac{1}{2u^{3}\ell^{5}}+\frac{1}{\varepsilon u}\left(\frac{1}{2\varepsilon^{2}}+\frac{1}{\ell^{2}}\right)\right].\label{eq:orbitequation}
\end{equation}

Note that the third term on the right gives the corrections to the
orbit due to the cosmological constant \cite{key-7} while the last
term gives the corrections due to non-commutative effects. For long
distances $\left(u\rightarrow0\right)$ or small noncommutative scale
$\left(\varepsilon\rightarrow0\right)$, the last term correctly tends
to zero giving the orbit analyzed in \cite{NCg5}.

\subsubsection{Advance of the Perihelion}

In order to obtain the advance of the perihelion of a planetary orbit
we use the method given in \cite{Corn} to compare a keplerian ellipse
in Lorentzian coordinates with one in noncommutative Schwarzschild-AdS
coordinates. The relevant relation communicating the two ellipse is
the constant of Kepler's second law. In Lorentz coordinates the line
element is given by

\begin{equation}
ds^{2}=-dt^{2}+dr^{2}+r^{2}(d\theta^{2}+\sin^{2}\theta d\phi^{2}).
\end{equation}

The noncommutative Schwarzschild-AdS gravitational field, given by
equation (\ref{eq:NCSAdS}), allow us to find the following transformation
of the coordinates, $r$ and $t$, in the binomial approximation

\begin{eqnarray}
dt' & = & \left(1-\frac{2M\gamma\left(\frac{3}{2};\frac{r^{2}}{4\varepsilon^{2}}\right)}{r\sqrt{\pi}}+\frac{r^{2}}{2L^{2}}\right)dt\\
dr' & = & \left(1+\frac{2M\gamma\left(\frac{3}{2};\frac{r^{2}}{4\varepsilon^{2}}\right)}{r\sqrt{\pi}}-\frac{r^{2}}{2L^{2}}\right)dr,
\end{eqnarray}

or using (\ref{eq:aux2}),

\begin{eqnarray}
dt' & = & \left(1-\frac{M}{r}-\frac{M}{\sqrt{\pi}\varepsilon}e^{-\frac{r^{2}}{4\varepsilon^{2}}}+\frac{r^{2}}{2L^{2}}\right)dt\label{eq:changet}\\
dr' & = & \left(1+\frac{M}{r}+\frac{M}{\sqrt{\pi}\varepsilon}e^{-\frac{r^{2}}{4\varepsilon^{2}}}-\frac{r^{2}}{2L^{2}}\right)dr.\label{eq:changer}
\end{eqnarray}

We consider two elliptical orbits, one the classical Kepler orbit
in $\left(r,t\right)$ space and a noncommutative Schwarzschild-AdS
orbit in $\left(r',t'\right)$ space. In the Lorentz space we have

\begin{equation}
dA=\int_{0}^{\rho}rdrd\phi,
\end{equation}
and hence the Kepler second law

\begin{equation}
\frac{dA}{dt}=\frac{1}{2}\rho^{2}\frac{d\phi}{dt}.
\end{equation}

In the noncommutative Schwarzschild-AdS situation, we have

\begin{equation}
dA'=\int_{0}^{\rho}rdr'd\phi.
\end{equation}

Therefore, using equation (\ref{eq:changer}) the integrand becomes

\begin{equation}
dA'=\int_{0}^{\rho}r\left(1+\frac{M}{r}-\frac{M}{\sqrt{\pi}\varepsilon}e^{-\frac{r^{2}}{4\varepsilon^{2}}}-\frac{r^{2}}{2L^{2}}\right)drd\phi
\end{equation}
which can be integrated to obtain

\begin{equation}
dA'=\frac{\rho^{2}}{2}\left(1+\frac{2M}{\rho}-\frac{\rho^{2}}{4L^{2}}+\frac{4M\varepsilon}{\sqrt{\pi}\rho^{2}}e^{-\frac{\rho^{2}}{4l^{2}}}-\frac{4M\varepsilon}{\sqrt{\pi}\rho^{2}}\right)d\phi.
\end{equation}
Hence, using (\ref{eq:changet}), the area law is

\begin{equation}
\frac{dA'}{dt'}=\frac{\rho^{2}}{2}\left(1+\frac{2M}{\rho}-\frac{\rho^{2}}{4L^{2}}+\frac{4M\varepsilon}{\sqrt{\pi}\rho^{2}}e^{-\frac{\rho^{2}}{4l^{2}}}-\frac{4M\varepsilon}{\sqrt{\pi}\rho^{2}}\right)\left(1-\frac{M}{\rho}-\frac{M}{\sqrt{\pi}l}e^{-\frac{\rho^{2}}{4\varepsilon^{2}}}+\frac{\rho^{2}}{2L^{2}}\right)^{-1}\frac{d\phi}{dt}
\end{equation}
and using again the binomial approximation,

\begin{eqnarray}
\frac{dA'}{dt'} & =\frac{\rho^{2}}{2} & \left[1+\frac{3M}{\rho}+\frac{2M^{2}}{\rho^{2}}-\frac{3\rho^{2}}{4L^{2}}-\frac{5M\rho}{4L^{2}}\right.\\
 &  & +\frac{2M\varepsilon}{\sqrt{\pi}}\left(\frac{1}{L^{2}}-\frac{2}{\rho^{2}}-\frac{2M}{\rho^{3}}\right)\left(1-e^{-\frac{\rho^{2}}{4\varepsilon^{2}}}\right)\\
 &  & \left.-\frac{M}{\sqrt{\pi}\varepsilon}e^{-\frac{\rho^{2}}{4l^{2}}}\left(1+\frac{2M}{\rho}-\frac{\rho^{2}}{4L^{2}}\right)+\frac{4M^{2}}{\pi\rho^{2}}e^{-\frac{\rho^{2}}{4\varepsilon^{2}}}\left(1-e^{-\frac{\rho^{2}}{4\varepsilon^{2}}}\right)+...\right]\frac{d\phi}{dt}.
\end{eqnarray}
Applying all of this increasing for a single orbit

\begin{eqnarray}
\Delta\phi' & =\int_{0}^{2\pi} & \left[1+\frac{3M}{\rho}+\frac{2M^{2}}{\rho^{2}}-\frac{3\rho^{2}}{4L^{2}}-\frac{5M\rho}{4L^{2}}\right.\\
 &  & +\frac{2M\varepsilon}{\sqrt{\pi}}\left(\frac{1}{L^{2}}-\frac{2}{\rho^{2}}-\frac{2M}{\rho^{3}}\right)\left(1-e^{-\frac{\rho^{2}}{4\varepsilon^{2}}}\right)\\
 &  & \left.-\frac{M}{\sqrt{\pi}\varepsilon}e^{-\frac{\rho^{2}}{4l^{2}}}\left(1+\frac{2M}{\rho}-\frac{\rho^{2}}{4L^{2}}\right)+\frac{4M^{2}}{\pi\rho^{2}}e^{-\frac{\rho^{2}}{4\varepsilon^{2}}}\left(1-e^{-\frac{\rho^{2}}{4\varepsilon^{2}}}\right)+...\right]d\phi.
\end{eqnarray}

For an ellipse (first approximation to the orbit), we have $\rho=\frac{R}{1+e\cos\phi}$,
where $e$ is the eccentricity and $R$ is the \emph{latus rectum}.
Applying the binomial approximation, we obtain

\begin{eqnarray}
\Delta\phi' & \approx & 2\pi+\frac{6\pi M}{\rho}+\frac{4\pi M^{2}}{\rho^{2}}-\frac{3\pi\rho^{2}}{2L^{2}}-\frac{5\pi M\rho}{2L^{2}}\\
 &  & +\frac{4\pi M\varepsilon}{\sqrt{\pi}}\left(\frac{1}{L^{2}}-\frac{2}{\rho^{2}}-\frac{2M}{\rho^{3}}\right)\left(1-e^{-\frac{\rho^{2}}{4\varepsilon^{2}}}\right)\\
 &  & -\frac{2\pi M}{\sqrt{\pi}\varepsilon}e^{-\frac{\rho^{2}}{4l^{2}}}\left(1+\frac{2M}{\rho}-\frac{\rho^{2}}{4L^{2}}\right)+\frac{4M^{2}}{\rho^{2}}e^{-\frac{\rho^{2}}{4\varepsilon^{2}}}\left(1-e^{-\frac{\rho^{2}}{4\varepsilon^{2}}}\right)+....
\end{eqnarray}

The classical advance of perihelion is recuperated for zero cosmological
constant (i.e. $L\rightarrow\infty$) and noncommutative limit $\left(\varepsilon\rightarrow0\right)$.
The last three terms are the corrections due to noncommutative geometry.

\subsubsection{Circular motion}

For circular motion in the equatorial plane we have $r=r_{c}=\mbox{constant}$
and so $\dot{r}=\ddot{r}=0$. The equation of the orbit (\ref{eq:orbitequation})
becomes

\begin{equation}
u_{c}=\frac{M}{\ell^{2}}+3Mu_{c}^{2}+\frac{1}{L^{2}\ell^{2}u_{c}^{3}}+\frac{M}{\sqrt{\pi}}e^{-1/4\varepsilon^{2}u_{c}^{2}}\left[\frac{3u_{c}}{\varepsilon}+\frac{1}{2u_{c}^{3}\ell^{5}}+\frac{1}{\varepsilon u_{c}}\left(\frac{1}{2\varepsilon^{2}}+\frac{1}{\ell^{2}}\right)\right]
\end{equation}
and the energy equation (\ref{eq:velocity}) is

\begin{equation}
E_{c}^{2}=V_{eff}^{2}\left(r_{c}\right)=\left(1-\frac{4M\gamma\left(\frac{3}{2};\frac{r_{c}^{2}}{4\varepsilon^{2}}\right)}{r_{c}\sqrt{\pi}}+\frac{r_{c}^{2}}{L^{2}}\right)\left(1+\frac{\ell^{2}}{r_{c}^{2}}\right).
\end{equation}

The radius of the circular orbit is determined by the condition

\begin{equation}
\left.\frac{d\left(V_{eff}^{2}\right)}{dr}\right|_{r=r_{c}}=0,
\end{equation}

that, using equation (\ref{eq:effectivepotential}) and (\ref{eq:auxdif}),
is 

\begin{equation}
\frac{4M\gamma\left(\frac{3}{2};\frac{r_{c}^{2}}{4\varepsilon^{2}}\right)}{r_{c}^{2}\sqrt{\pi}}\left[1+\frac{3\ell^{2}}{r_{c}^{2}}\right]-\frac{M}{\sqrt{\pi}\varepsilon^{3}}e^{-\frac{r_{c}^{2}}{4\varepsilon^{2}}}\left[r_{c}+\frac{\ell^{2}}{r_{c}}\right]+\frac{2r_{c}}{L^{2}}-\frac{2\ell^{2}}{r_{c}^{3}}=0.\label{eq:aux4}
\end{equation}
The stability of the circular orbit is given by 

\begin{equation}
\left.\frac{d^{2}\left(V_{eff}^{2}\right)}{dr^{2}}\right|_{r=r_{c}}\geq0
\end{equation}
or

\begin{equation}
-\frac{8M\gamma\left(\frac{3}{2};\frac{r_{c}^{2}}{4\varepsilon^{2}}\right)}{r_{c}^{3}\sqrt{\pi}}\left[1+\frac{6\ell^{2}}{r_{c}^{2}}\right]+\frac{M}{\sqrt{\pi}\varepsilon^{3}}e^{-\frac{r_{c}^{2}}{4\varepsilon^{2}}}\left[\frac{4\ell^{2}}{r_{c}^{2}}+\frac{r_{c}^{2}}{2\varepsilon^{2}}+\frac{\ell^{2}}{2\varepsilon^{2}}\right]+\frac{2}{L^{2}}+\frac{6\ell^{2}}{r_{c}^{4}}\geq0.\label{eq:aux5}
\end{equation}

Combining equations (\ref{eq:aux4}) and (\ref{eq:aux5}), we obtain
the condition

\begin{equation}
\frac{8\ell^{2}}{r_{c}^{4}}-\left[\frac{5\ell^{2}}{r_{c}^{2}}+1\right]\frac{12M\gamma\left(\frac{3}{2};\frac{r_{c}^{2}}{4\varepsilon^{2}}\right)}{r_{c}^{3}\sqrt{\pi}}+\left[r_{c}^{2}+\frac{10\ell^{2}\varepsilon^{2}}{r_{c}^{2}}+\ell^{2}+2h\varepsilon^{2}\right]\frac{M}{2\sqrt{\pi}\varepsilon^{5}}e^{-\frac{r_{c}^{2}}{4\varepsilon^{2}}}\geq0.
\end{equation}

This is a complicated relation with no analytical solution for $r_{c}$.
Instead, we have depicted the left hand side of this relation in terms
of the radius. The result is shown in Figure 6 and compared with the
Schwarzschild spacetime. Note that in commutative Schwarzschild geometry,
the circular orbits are stable when $r_{c}\geq6M$ \cite{Carroll},
while in the noncommutative Schwarzschild-AdS spacetime the ciruclar
orbits are always stable.

\begin{figure}
\begin{centering}
\includegraphics[scale=0.25]{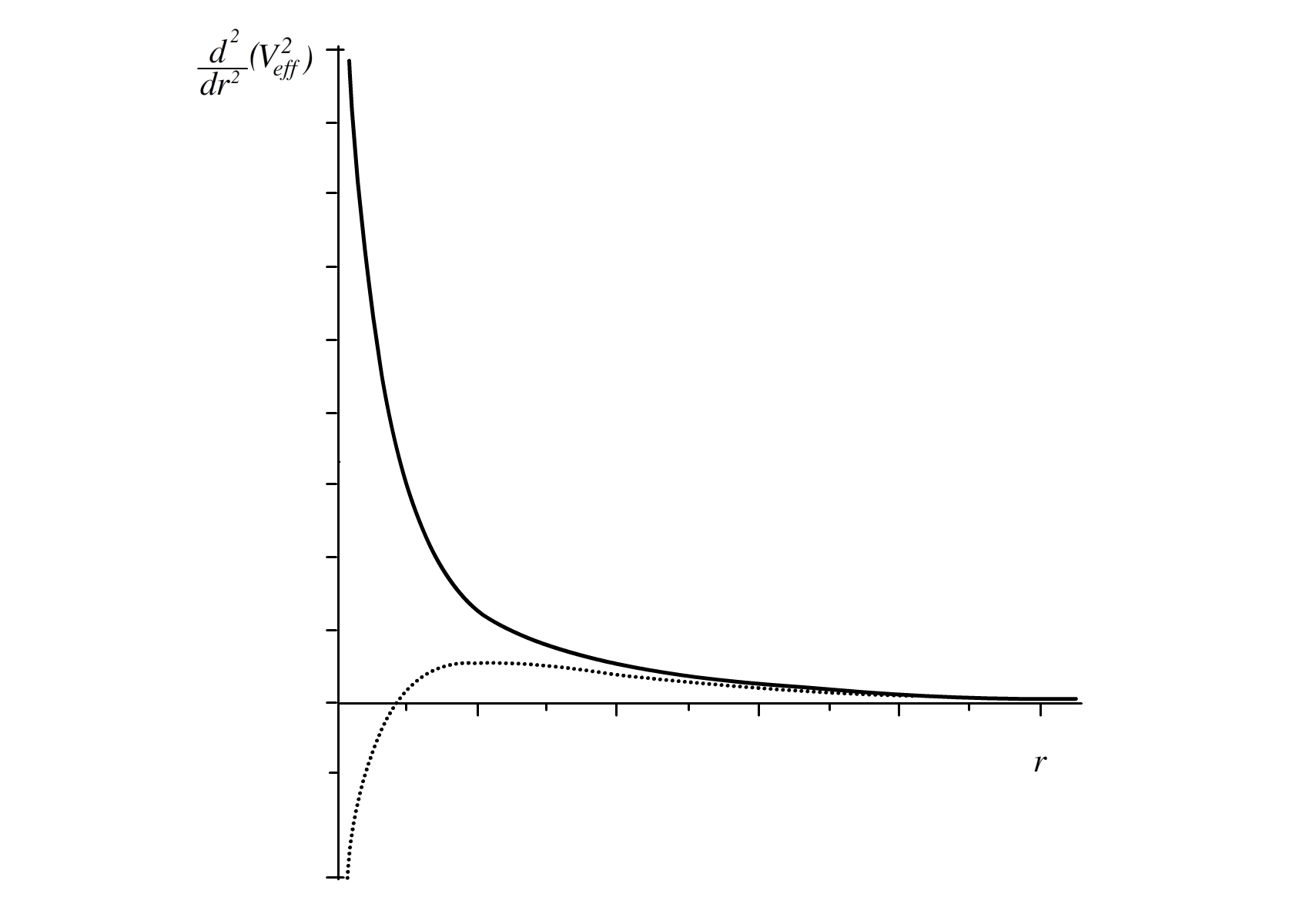}
\par\end{centering}

\caption{The condition for stability of circular orbits of particles in Schwarzschild
spacetime (dotted curve) and noncommutative Schwarzschild-AdS (continuous
curve). In the commutative case the condition for stability is given
by $r\geq6GM$. In the noncommutative situation the circular orbits
are always stable. }
\end{figure}

\section{Conclusion}

In this paper we have studied the effects of noncommutativity in the
orbits of particles in Schwarzschild-AdS spacetime. By means of a
detailed analysis of the corresponding effective potentials for particles,
we find all possible motions which are allowed by the energy levels.
For radial time-like geodesics, there are some bounded trajectories
(depending on the exact values of the parameters). Therefore particles
not always plunges into $r=0$ from an upper distance.

For non-radial time-like geodesics, elliptical orbits are allowed
as well as circular orbits. We also calculated the effect of space
noncommutativity on the value of the precession of the perihelion,
giving an infinity serial including the cosmological constant contribution
reported in \cite{key-7} and the noncommutative terms. Although this
noncommutative effect is very small, it is important since reflect
the nature of spacetime structure at quantum gravity level. Therefore,
the geodesic structure of this black hole presents new types of motion
not allowed by the Schwarzschild spacetime. Finally, the stability
of circular orbits in noncommutative Schwarzschild-AdS spacetime is
discussed, showing a new behavior when compared with the commutative
Schwarzchild case.

In a forthcoming paper we will discuss the null geodesic structure
of the noncommutative Schwarzschild-AdS spacetime.\\

\emph{Acknowledgement}

This work was supported by the Universidad Nacional de Colombia. Hermes
Project Code 13038.

\end{document}